\documentclass[titlepage,11pt,twoside]{article}\usepackage[]{graphicx}\usepackage[]{color}
\makeatletter
\def\maxwidth{ %
  \ifdim\Gin@nat@width>\linewidth
    \linewidth
  \else
    \Gin@nat@width
  \fi
}
\makeatother

\definecolor{fgcolor}{rgb}{0.345, 0.345, 0.345}

\usepackage{framed}
\makeatletter
 {\par\unskip\endMakeFramed%
 \at@end@of@kframe}
\makeatother

\definecolor{shadecolor}{rgb}{.97, .97, .97}
\definecolor{messagecolor}{rgb}{0, 0, 0}
\definecolor{warningcolor}{rgb}{1, 0, 1}
\definecolor{errorcolor}{rgb}{1, 0, 0}
\usepackage{alltt}
\usepackage{natbib}
\usepackage{amsmath,amsthm,amssymb,enumerate,enumitem}
\usepackage{mathtools}
\usepackage{booktabs} % for \midrule and \cmidrule macros
\usepackage[svgnames]{xcolor}
\usepackage{multirow}
\usepackage{tabularx}
\usepackage{listings}
\usepackage[myheadings]{fullpage}
\usepackage{pmetrika}
\usepackage{amsmath,amssymb,url}

\newcommand{\E}{\mathbb{E}}

\newcommand{\R}{\mathbb{R}}

\IfFileExists{upquote.sty}{\usepackage{upquote}}{}
\begin{document}

\begin{titlepage}

\title{Accurate confidence interval estimation for non-centrality parameters and effect size indices}
\author{Kaidi Kang$^1$; Kristan Armstrong$^2$; Suzanne Avery$^2$; Maureen McHugo$^2$; Stephan Heckers$^2$; Simon Vandekar$^1$}

\affil{$^1$Vanderbilt University, Department of Biostatistics}
\affil{$^2$Vanderbilt University Medical Center, Department of Psychiatry and Behavioral Sciences}

~\\
Please address correspondence to: \\
Kaidi Kang\\
2525 West End Ave., \#1136\\
Department of Biostatistics\\
Vanderbilt University\\
Nashville, TN 37203\\
{\tt kaidi.kang@vanderbilt.edu}

\linespacing{1}

\begin{center}\vskip3pt

\newpage

Abstract\vskip3pt

\end{center}
\begin{abstract}
% Background
%Effect size indices are useful tools for communicating study findings because they are unitless and do not depend on the sample size.
%Reporting effect size index estimates with their confidence intervals can be an excellent way to simultaneously communicate the strength of the observed evidence and the confidence one should have in that evidence.
We recently proposed a robust effect size index (RESI) that is related to the non-centrality parameter of a test statistic.
RESI is advantageous over common indices because (1) it is widely applicable to many types of data; (2) it can rely on a robust covariance estimate; (3) it can accommodate the existence of nuisance parameters. 
We provided a consistent estimator for the RESI, however, there is no established confidence interval (CI) estimation procedure for the RESI.
Here, we use statistical theory and simulations to evaluate several CI estimation procedures for three estimators of the RESI.
%We evaluate the influence of homo-/hetero-skedasticity, data skewness and fixed/random covariates on the performance of the confidence intervals.
% Results
Our findings show (1) in contrast to common effect sizes, the robust estimator is consistent for the true effect size; %(2) counter to intuition, the randomness of covariates reduces coverage for Chi-squared and F CIs so they are not effective in practice; (3) When the variance of the parameter estimators is estimated, then non-central Chi-squared and F CIs using the parametric and robust RESI estimators
(2) common CI procedures for effect sizes that are non-centrality parameters fail to cover the true effect size at the nominal level.
Using the robust estimator along with the proposed bootstrap CI is generally accurate and applicable to conduct consistent estimation and valid inference for the RESI, especially when model assumptions may be violated.
Based on the RESI, we propose a general framework for the analysis of effect size (ANOES), such that effect sizes and confidence intervals can be easily reported in an analysis of variance (ANOVA) table format for a wide range of models.

\end{abstract}


\vspace{\fill}
\end{titlepage}\vspace*{24pt}

% {$^\text{a}$Department of Biostatistics and Epidemiology, University of Pennsylvania, Philadelphia PA 19104, USA}
% {$^\text{b}$Department of Psychiatry, University of Pennsylvania, Philadelphia PA 19104, USA}
% {$^\text{c}$Philadelphia Veterans Administration Medical Center, Philadelphia PA 19104, USA}
% {$^\text{d}$Department of Radiology, University of Pennsylvania, Philadelphia PA 19104, USA}

% STUFF TO SETUP FIGURES AND R OPTIONS

\section{Introduction}
% what are effect sizes
Effect size indices are measures quantifying the strength of association between a covariate and an outcome of interest that are unaffected by the sample size. They play an important role in power analyses, sample size planning and meta-analyses \citep{cohen_statistical_1988, chinn_simple_2000, morris_combining_2002}. 
% why ppl should be interested in effect sizes;
Given the recent criticism of the misuse and misinterpretation of null hypothesis significance testing by the American Statistical Association (ASA) \citep{wasserstein_asas_2016,wasserstein_moving_2019}, there is an increasing interest in seeking alternatives for communicating study findings. 
Reporting effect size estimates with confidence intervals (CIs) can be an excellent way to simultaneously communicate the strength of evidence (sample size independent) as well as the confidence one should have in the evidence (sample size dependent).
The American Psychological Association (APA) has also been calling for reporting of effect sizes and their CIs for almost two decades \citep{american_psychological_association_publication_2001, american_psychological_association_publication_2010}, but they are still not routinely reported by psychological studies. \cite{fritz_effect_2012} reviewed articles published in 2009 and 2010 in the \emph{Journal of Experimental Psychology: General}, and noted that less than half of the articles they reviewed reported effect sizes and no article reported a confidence interval for an effect size. 
The barriers stopping researchers from easily reporting effect sizes along with their CIs not only lies in their unfamiliarity of different effect size indices but also in the lack of guidance of how to correctly estimate the CI for a specific effect size index. 

% RESI
We recently proposed a robust effect size index (RESI) \citep{vandekar_robust_2020}, which has several advantages over previously proposed indices \citep{cohen_statistical_1988, hedges_statistical_1985, long_regression_2006,zhang_robust_1997,rosenthal_parametric_1994} because (1) it is widely applicable to many types of data since it is constructed from M-estimators, which are generally defined; (2) it can rely on a robust covariance estimate; (3) it can accommodate the existence of nuisance covariates/parameters.
We also proposed a simple consistent estimator for the RESI that is a function of the Chi-squared test statistic \citep{vandekar_robust_2020}.
The RESI is defined on the basis of the Wald test statistic -- it is related to the non-centrality parameter (NCP) of the test statistic under the alternative hypothesis, therefore, 
% there is a parametric version (which uses the theoretical covariance estimator) and a robust version, which relies on the robust ``sandwich" covariance estimator \citep{white_heteroskedasticity-consistent_1980,long_using_2000,huber_robust_1964,mackinnon_heteroskedasticity-consistent_1985}.
it has the generality that it can be estimated in different types of data.
Researchers can use the RESI to report their observed effect sizes regardless of the model.
Furthermore, the studies on the same scientific topic, but using different types of data can easily communicate their observed effect sizes without translating between different effect size indices.
While the RESI can currently be used to report the strength of a finding through the RESI estimate, we did not establish a CI estimation procedure for the RESI, which quantifies the amount of certainty of the estimate.

% Objective(s) of this paper
The goal of this paper is to establish an accurate CI estimation procedure for the RESI and establish a framework for the analysis of effect sizes (ANOES) based on the RESI.
Because the RESI is related to the NCP of the Chi-squared statistic, an intuitive approach is to use existing methods of constructing intervals for the NCP of a noncentral Chi-squared or F distribution \citep{kent_confidence_1995, harlow_what_2013}.
Here, we use statistical theory and simulations to show that the Chi-squared CI provides low coverage for the NCP when the variance must be estimated.
In fact, the coverage gets lower with increasing sample and effect size.
Similarly, the F CI has decreasing coverage with increasing sample size when the robust covariance estimate is used.
We use theory to show that this occurs because the variance of the estimators is not asymptotically equivalent.
As a solution, we propose several bootstrap CI construction procedures and evaluate their coverage performance through simulations in various scenarios.
We propose a bootstrap CI for the robust estimator of the RESI that is generally accurate and applicable, even when model assumptions are violated. 
Based on the RESI, we propose a general framework for ANOES, such that effect sizes and CIs can be easily reported in an analysis of variance (ANOVA) table format. 
We use this framework to study the effect of early psychosis and schizophrenia on relational memory for illustrative purpose. Our early stage {\tt RESI R} package is available to install through github (\url{https://github.com/statimagcoll/RESI}).

\newpage

\section{Statistical theory}

\subsection{Estimators for the Robust Effect Size Index (RESI)} \label{s:estimators}
In this section, we define the RESI and describe three estimators for the parameter.
Let $W = (W_1, \ldots, W_n)$ denote the full dataset, where $W_i$ is independent of $W_j$ for all $i\ne j$.
Assume $\theta = (\alpha, \beta)  \in \Theta \subset \R^m$ are model parameters, where $\alpha \in \R^{m_0}$ is a vector of nuisance parameters, $\beta \in \R^{m_1}$ is a vector of target parameter, and $m = m_0 + m_1$.
We assume $\Psi(\theta; W) =  n^{-1}\sum_{i=1}^n \psi(\theta; W) \in R$ is an estimating equation, where $\psi$ is a known function and $\Psi$ can be maximized to obtain the M-estimator $\hat{\theta}$
\begin{equation*}
    \hat{\theta} = \operatorname*{arg\,max}_{\theta^* \in \Theta} \Psi(\theta^*; W).
\end{equation*}
If $\Psi$ is a likelihood function then $\hat\theta$ corresponds to the maximum likelihood estimator.
Denote the target value of the parameter $\theta = \arg \max_{\theta^*}\E \Psi(\theta^*; W)$.

We define the components of the asymptotic covariance of $\sqrt{n}(\hat{\theta}-\theta)$:
\begin{equation*}
    \mathbf{J}_{jk}(\theta) = - \left. \lim_{n\to\infty}\mathbb{E} \frac{\partial^2 \Psi(\theta^*, W)}{\partial \theta^*_j \partial \theta^*_k} \right\rvert _\theta 
\end{equation*}
\begin{equation*}
    \mathbf{K}_{jk}(\theta) = \left. \lim_{n\to\infty} \E \frac{\partial \Psi(\theta^*, W)}{\partial \theta^*_j} \frac{\partial \Psi(\theta^*, W)}{\partial \theta^*_k} \right \rvert_\theta
\end{equation*}

Under mild conditions \citep{van_der_vaart_asymptotic_2000,vandekar_robust_2020}, the variance of $\sqrt{n}(\hat{\theta}-\theta)$ is
\begin{equation}\label{eq:sigmaTheta}
    \Sigma_{\theta} = \mathbf{J}(\theta)^{-1} \mathbf{K}(\theta) \mathbf{J}(\theta)^{-1}
\end{equation}
If $\Psi$ is a correctly specified likelihood function, then $\mathbf{J}(\theta) = \mathbf{K}(\theta)$, and the asymptotic covariance matrix of $\sqrt{n}(\hat{\theta}-\theta)$ is $\mathbf{J}(\theta)^{-1}$.

We defined the RESI from the test statistic for $H_0: \beta = \beta_0$, where $\beta_0$ is a vector-valued reference point.
We, previously, suggested that the typical Wald-style statistic for the test of the null hypothesis follows a Chi-squared distribution with $m_1$ degrees of freedom and non-centrality parameter $n(\beta - \beta_0 )^T \Sigma_{\beta} (\beta - \beta_0)$,
\begin{align} \label{eq:T}
    T^2 = n(\hat{\beta} - \beta_0 )^T \Sigma_{\beta}^{-1}(\hat\theta) (\hat{\beta} - \beta_0) \sim \chi^2_{m_1}\{ n(\beta - \beta_0 )^T \Sigma_{\beta}^{-1}(\theta) (\beta - \beta_0) \},
\end{align}
where the $\hat{\beta}$ is the estimated value of $\beta$ and $\Sigma_{\beta}(\theta)$ is the asymptotic covariance matrix of $\sqrt{n}(\hat{\beta}-\beta)$.

The RESI was defined as the square root of the component of the NCP that is due to the deviation of $\beta$ from the reference value,
\begin{align*}
    S_{\beta} = \sqrt{(\beta - \beta_0 )^T \Sigma_{\beta}^{-1}(\theta) (\beta - \beta_0)}.
\end{align*}

The estimator for the RESI is defined as \citep{vandekar_robust_2020}
\begin{equation} \label{eq:S_hat}
    \hat{S}_{\beta} = \left\{\max\left[0, \frac{T^2 - m_1}{n} \right] \right\}^{\frac{1}{2}}
\end{equation}
where $T^2$ is as defined in \eqref{eq:T}.
%This estimator was derived via a method similar to the method of moments: the statistic $T_{m_1}^2$ is known to follow a non-central Chi-squared distribution $\chi^2_{m_1}\{ n S_{\beta}^2 \}$.
This estimator was derived by setting the observed statistic $T^2$ equal to the expected value of the non-central Chi-squared distribution and solving for $S_\beta$
\begin{align*}
    \hat{S}_{\beta}^2 & = \frac{T^2 - m_1}{n}.
\end{align*}
Because $S_{\beta}$ must be nonnegative, the estimator \eqref{eq:S_hat} has lower mean square error \citep{vandekar_robust_2020,neff_further_1976,kubokawa_estimation_1993}.

There are several ways of constructing the statistic $T^2$ and $\hat S_\beta$ through the estimation of $\Sigma_{\beta}(\hat\theta)$ \citep{white_heteroskedasticity-consistent_1980,mackinnon_heteroskedasticity-consistent_1985,long_using_2000}.
First, the matrices $\mathbf{J}$ and $\mathbf{K}$ are estimated by
\begin{align}
\hat{\mathbf{J}}_{jk}(\hat\theta) & = - \left. n^{-1}\sum_{i=1}^n  \frac{\partial^2 \psi(\theta^*, W)}{\partial \theta^*_j \partial \theta^*_k}  \right\rvert _{\hat\theta} \label{eq:Jestimate}\\
\hat{\mathbf{K}}(\hat\theta) & = \left. n^{-1}\sum_{i=1}^n \frac{\partial \psi(\theta^*, W)}{\partial \theta^*_j} \frac{\partial \psi(\theta^*, W)}{\partial \theta^*_k} \right\rvert_{\hat{\theta}} \notag \\
\hat{\Sigma}(\hat\theta) &= \hat{\mathbf{J}}(\hat\theta)^{-1} \hat{\mathbf{K}}(\hat\theta) \hat{\mathbf{J}}(\hat\theta)^{-1} \label{eq:sigmaHat}.
\end{align}
%Since the RESI was defined to be non-negative, if we force the estimator to be non-negative and replace the $n$ in the denominator with the residual degree(s) of freedom to account for the estimation of $m_1$ parameters, we will have the estimator in \eqref{eq:S_hat}.\\
Then we can estimate the RESI using three versions of test statistics: 
\begin{enumerate}
    \item \textbf{Oracle test statistics}: when the true covariance of $\sqrt{n}(\hat{\beta}-\beta)$ is known, 
    \begin{equation*}
    T_{(o)}^2 = n(\hat{\beta} - \beta_0 )^T \Sigma_{\beta}^{-1}(\theta) (\hat{\beta} - \beta_0),
    \end{equation*}
    where $\Sigma_{\beta}(\theta)$ is obtained from the block diagonal of \eqref{eq:sigmaTheta} corresponding to $\beta$. This is called an ``oracle" statistic because it depends on the true covariance matrix, which is not known in practice.
    \item \textbf{Parametric test statistics}: when we believe the working model is correctly specified,
    \begin{equation*}
    T_{(p)}^2 = n(\hat{\beta} - \beta_0 )^T \hat{\mathbf{J}}_{\beta}(\hat\theta) (\hat{\beta} - \beta_0),
    \end{equation*}
    where $\hat{\mathbf{J}}_{\beta}(\hat\theta)$ is obtained from the block diagonal component of \eqref{eq:Jestimate} corresponding to $\beta$.
    \item \textbf{Robust test statistics:} when the working model is not assumed to be correct,
    \begin{equation*}
    T_{(r)}^2 = n(\hat{\beta} - \beta_0)^T \hat\Sigma_{\beta}^{-1}(\hat\theta) (\hat{\beta} - \beta_0),
    \end{equation*}
    where $\hat\Sigma_{\beta}(\hat\theta)$ is obtained from the block diagonal component of \eqref{eq:sigmaHat} corresponding to $\beta$.
\end{enumerate}

With the oracle, parametric, or robust test statistics, we can derive the corresponding oracle, parametric or robust estimator for the RESI using \eqref{eq:S_hat}.
These different versions of the estimator for $S_\beta$ have different sampling distributions depending on which test statistic is used.

\subsection{Chi-squared CIs underestimate the variance} \label{sec: linear_case}
% The variance of the test stat depends on diff test statistics under the alternative
% the var of test stats under the alternative are not asymptotically equal
When the covariance of $\sqrt{n}(\hat{\beta}-\beta)$ is known, the test statistics follows the non-central Chi-squared distribution \eqref{eq:T}, where its variance depends on the true effect size.
However, in practice, the covariance needs to be estimated.
% and (we will show that) the variance of the test statistic is increased.
In this section, we will show that the large sample distribution of the test statistic deviates from the theoretical non-central Chi-squared distribution when an estimator for the covariance is used.
As a result, the CIs constructed using Chi-squared distribution \citep{kent_confidence_1995} will fail to provide nominal coverage.
This is in contrast to the null case, where the asymptotic distribution of the Chi-square statistic is valid whether or not $\Sigma_{\beta}(\theta)$ is known or estimated.
To illustrate this problem here, we compare the asymptotic equivalence of the variance of the oracle, parametric, and robust estimators in a linear model.
Two functions $f(n)$, $g(n)$ are said to be asymptotically equivalent if $\lim_{n \to \infty} \frac{f(n)}{g(n)} = 1$.
We need asymptotic equivalence here, because a standard asymptotic approach cannot be used; under the alternative, the mean and variance of the test statistics depend on the sample size.

Throughout this section, we assume $Y = X\beta + \epsilon$, where $X$ is full rank and $\epsilon_i$ are independent with mean 0 and variance $\sigma^2_i$. Note that $\sigma^2_i$ may not be equal.
The ordinary least squares (OLS) estimator of $\beta$ is $\hat{\beta}_{OLS} = (X^T X)^{-1} X^T Y$. 

\paragraph{Oracle Estimator: } Assuming known covariance, the test statistic \eqref{eq:T} is approximately central Chi-squared by the central limit theorem under the null.
Under the alternative, a similar argument suggests its approximation to a non-central Chi-squared distribution.
\begin{equation*}
    T_{(o)}^2 \sim \chi^2_{m_1}( nS^2_{\beta} ).
\end{equation*}
%In practice, the covariance of the parameter estimator is rarely known but results under the null suggest plugging in the estimated variance and using the same approximation.
%We will show, however, that this leads to confidence intervals with less than nominal coverage because the asymptotic distribution is not accurate.

The expectation and variance of the oracle test statistics are:
\begin{align} \label{eq: var_oracle_est}
    \E T_{(o)}^2 & = m_1 + nS^2_{\beta} \notag \\
    Var(T_{(o)}^2) & = 2(m_1 + 2nS^2_{\beta}).
\end{align}
Thus, the estimator \eqref{eq:S_hat} is consistent and the variance of $T_{(o)}$ is linear in the sample size.

\paragraph{Parametric Estimator: }
Assuming homoskedasticity (i.e., $\sigma^2_i = \sigma^2$, $\forall i$) and normality of $Y$ makes finding the distribution of the test statistic tractable.
The covariance matrix of $\hat{\beta}_{OLS}$ can be estimated as $ (X^T X)^{-1} \hat{\sigma}^2$, where $\hat{\sigma}^2 = (n-m)^{-1} Y^T (I - H) Y$ and $H$ is the hat matrix $H = X (X^TX)^{-1} X^T$.
In this situation, the parametric version of test statistic for $H_0:\beta = \beta_0$ is
\begin{align} \label{eq: para_test_stat}
        T_{(p)}^2 
        & = n (\hat{\beta}_{OLS} - \beta_0)^T \hat{\Sigma}_{\beta_{OLS}}^{-1} (\hat{\beta}_{OLS} - \beta_0) \notag \\
        & = \hat{\sigma}^{-2} (\hat{\beta}_{OLS} - \beta_0)^T X^T X (\hat{\beta}_{OLS} - \beta_0) 
\end{align}

%If the covariance doesn't need to be estimated, the test statistics in \eqref{eq:T} follow non-central Chi-squared distributions.
Additional variability is introduced into the test statistic when plugging in the estimated covariance of $\hat{\beta}_{OLS}$ and, consequently, the parametric test statistic does not follow the non-central Chi-squared distribution \eqref{eq:T}.
For linear regression models, it can be shown that the test statistics divided by its own degrees of freedom $m_1$ follows the non-central F-distributions (see Appendix)
\begin{equation} \label{eq:F_dist}
T_{(p)}^2 / m_1 \sim F(m_1, n - m; nS^2_{\beta}).
\end{equation}

By plugging the parametric test statistic \eqref{eq: para_test_stat} into \eqref{eq:S_hat}, we can derive a parametric version of the estimator for the RESI.
The expectation and variance of the parametric test statistic are derived from the moments of the F-distribution,
\begin{align*}
     \E T_{(p)}^2 & = \frac{(n-m)(m_1 + nS^2_{\beta})}{n-m - 2}\\
     Var\left(T_{(p)}^2 \right) & = 2 \frac{(m_1 + nS^2_{\beta})^2 + (m_1 + 2nS^2_{\beta})(n-m-2)}{(n-m-2)^2 (n-m-4)}\left(n-m \right)^2
\end{align*}
$\E T_{(p)}^2$ and $\mathrm{Var} \left(T_{(p)}^2 \right)$ are asymptotically equivalent to
\begin{align}
    \E T_{(p)}^2  &\simeq m_1 + nS^2_{\beta}\notag\\
    \mathrm{Var} \left(T_{(p)}^2 \right) &\simeq 2(m_1 + 2nS^2_{\beta}) + 2\left( 2 m_1 S^2_{\beta} + nS^4_{\beta} \right)\label{eq:var_param}\\
    & = \mathrm{Var} \left(T_{(o)}^2 \right) + 2\left( 2 m_1 S^2_{\beta} + nS^4_{\beta} \right).\notag
\end{align}
This parametric test statistic, which is simplest unknown variance case, has a variance that is not asymptotically equal to the variance of the oracle test statistic, which follows the non-central Chi-squared distribution.
The consequence of this fact is that a CI using the Chi-squared distribution will have lower than nominal coverage for the NCP and the effect sizes that are a function of the NCP.
In fact, the CI will become increasingly inaccurate as the effect size or sample size get larger.
We demonstrate this with simulations in Section \ref{sec: sim_1}. 

\paragraph{Robust Estimator: }
When there is a suspected unknown heteroskedasticity, a robust version of covariance estimator of $\hat{\beta}_{OLS}$ can be applied instead,
\begin{align*}
    \hat{\mathrm{V}}\mathrm{ar}(\hat{\beta}_{OLS}) = (X^T X)^{-1}X^T \hat{G} X (X^T X)^{-1}
\end{align*}
where $\hat{G}_{ij} = \frac{e_i^2}{(1 - h_i)^2}$ if $i = j$ and $\hat{G}_{ij} = 0$ if $i \ne j$, $e_i = Y_i - X_i\hat{\beta}$ is the $i$th residual, and $h_i$ is the $i$-th elements on the diagonal of the hat matrix $H$, i.e., $h_i = x_i (X^TX)^{-1} x_i^T$ \citep{long_using_2000}.
Many versions of this robust ``sandwich" covariance estimator have been proposed, this version is discussed by \citet{long_using_2000} as a jackknife approximation.

The robust version of the test statistic in this case is
\begin{align} \label{eq: robust_test_stat}
        T_{(r)}^2 
        & = n (\hat{\beta}_{OLS} - \beta_0)^T \hat{\Sigma}_{\beta}^{-1}(\hat{\theta}) (\hat{\beta}_{OLS} - \beta_0) \notag  \\
        & = (\hat{\beta}_{OLS} - \beta_0)^T (X^T X) (X^T \hat{G} X)^{-1} (X^T X) (\hat{\beta}_{OLS} - \beta_0) .
\end{align}
Then the robust version of the estimator for RESI can be constructed by plugging \eqref{eq: robust_test_stat} into \eqref{eq:S_hat}.

Similar to the situation of parametric test statistic, the covariance of parameter estimator $\hat{\beta}_{OLS}$ is estimated.
The estimator is still consistent
\begin{align*}
\E T_{(r)}^2 & = \mathrm{tr} \left( \left( n^{-1} \hat{\Sigma}_{\hat{\beta}_{OLS}} \right)^{-1} n^{-1}\Sigma_{\hat{\beta}_{OLS}} \right) + n(\E \hat{\beta}_{OLS} - \beta_0)^T  \hat{\Sigma}_{\hat{\beta}_{OLS}}^{-1} (\E \hat{\beta}_{OLS} - \beta_0) \\
& \simeq \mathrm{tr} \left( \Sigma_{\hat{\beta}_{OLS}}^{-1} \Sigma_{\hat{\beta}_{OLS}} \right) + n(\beta - \beta_0)^T  \Sigma_{\hat{\beta}_{OLS}}^{-1} (\beta - \beta_0) \\
& = \mathrm{tr} \left( \mathbf{I}_{m_1} \right) + nS^2 \\
& = m_1 + nS^2 = \E T^2_{(o)},
\end{align*} 
but using the robust covariance estimator increases variance of the robust test statistic above the non-central Chi-squared and the F distributions.
We found an analytical form for the variance of $T_{(r)}^2$ to be intractable, so estimated it using simulations.

\begin{figure}[h]
    \centering
    \includegraphics[width = 15cm]{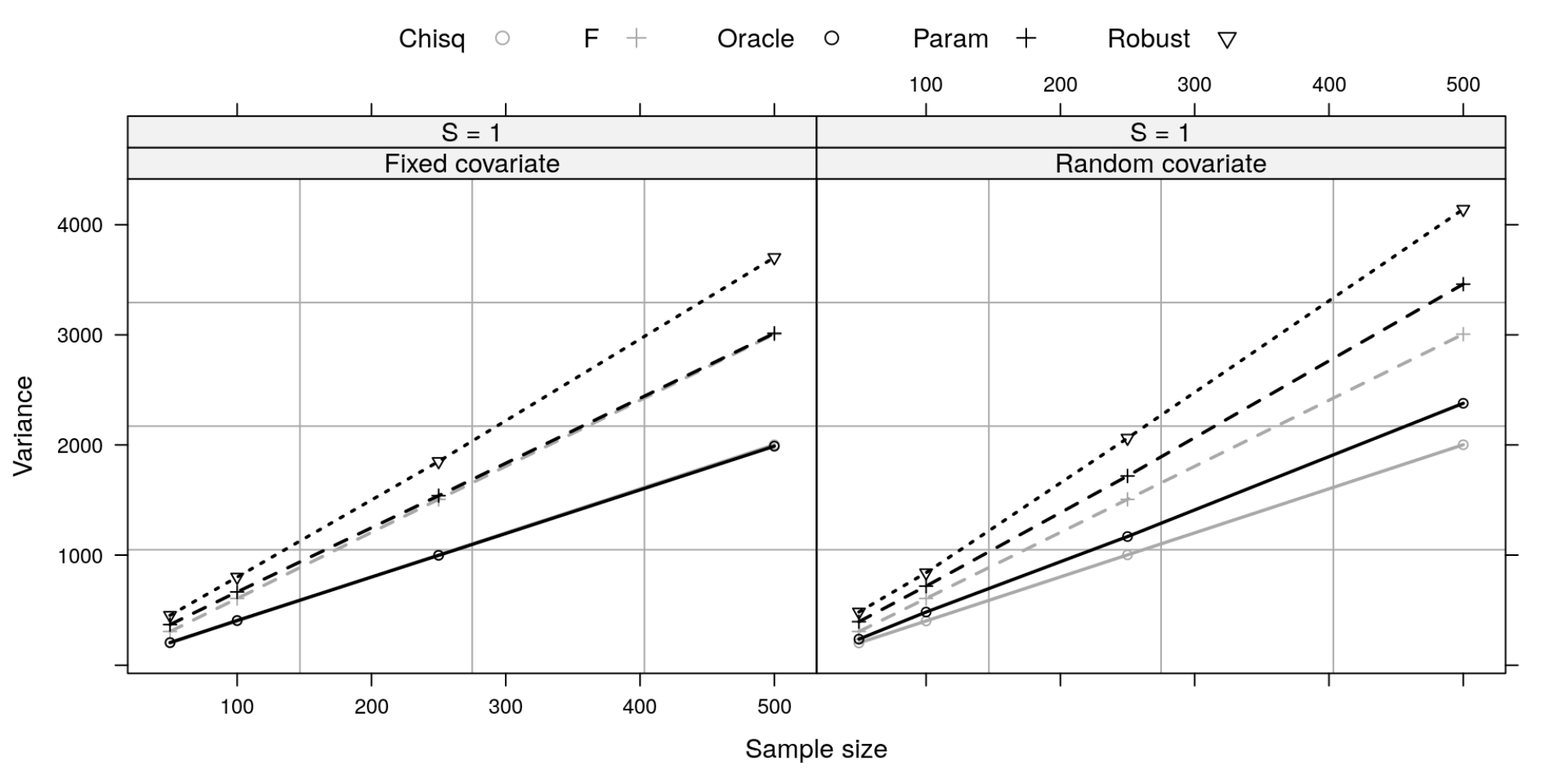}
    \caption{A comparison of simulated variance of the three test statistics from a simple linear regression model with a binary covariate and assumptions of homoskedasticity and symmetric errors. 
    The grey lines are the variance of theoretical non-central F or Chi-squared distribution using formulas \eqref{eq: var_oracle_est} and \eqref{eq:var_param}. The black lines are the simulated variance of test statistics with random or fixed covariate.
    With fixed covariate, the variance of oracle and parametric statistics equals theoretical variance of the non-central Chi-squared or F distribution, respectively.
    With random covariates, the true variance of the test statistic deviates from the theoretical value.
    $S$ denotes the RESI, Chisq and F denote the non-central Chi-squared and F distributions, respectively.
    }
    \label{fig: var_of_stat}
\end{figure}

Figure \ref{fig: var_of_stat} shows the simulated variance of each test statistic as a linear function of $n$.
When the effect size is large (e.g., $S = 1$) and with a fixed covariate, it's obvious that the variance of oracle and parametric statistics equals the variance of non-central Chi-squared and F distributions, respectively. This is expected since these two distributions are the asymptotic distributions of these two statistics, respectively (Section \ref{sec: linear_case}).
Because the test statistic using the robust covariance has larger variance than the oracle and parametric statistics, the robust test statistic is not non-central Chi-squared or non-central F distributed. This implies that the CIs constructed using either of these two distributions will not have accurate coverage for the robust RESI estimator and that the coverage will get worse with increasing effect size or sample size.
When the covariate is random, the variance of all three statistics further grows, and both of the oracle and parametric statistics deviate from the non-central Chi-squared and F distributions. 
This implies that when the covariate(s) is random by designs (such as observational studies), neither of these two distributions will provide CIs with accurate coverage for either of these three estimators. 
While we only studied linear models in this section, we expect that the robust test statistic will not follow the Chi-squared or F distribution under the alternative in general.

% The variance of the three test statistics is not asymptotically equal. The variance of the robust test statistic is always larger than the variance of oracle test statistic, which follows the non-central Chi-squared distribution. The variance of the parametric test statistic is always larger than the variance of parametric test statistic, which follows the non-central F distribution (Figure \ref{fig: var_of_stat}). This implies that, when using a robust estimator for the covariance, the robust test statistic doesn't follow the Chi-squared distribution (as the oracle test statistic) nor the F distribution (as the parametric test statistic). 
%The variance of test statistics depends on the sample size and true effect size. The difference between the variance of the three test statistics increases as the effect size or sample size gets larger, as we expected and discussed in Section \ref{sec: linear_case}.

\section{Methods}

\subsection{Confidence Intervals for RESI}

Here, we discuss several potential procedures to construct CIs using non-central Chi-squared and F distributions and bootstrapping.\\

\subsubsection{Theoretical methods: using non-central distributions\\} 
As discussed in the previous section, the test statistic may follow a non-central Chi-squared or F distribution with a non-centrality parameter (NCP) not equal to 0.
Since the squared RESI was defined as the NCP divided by sample size $n$, there is a one-to-one relationship between the NCP and RESI.
If a CI for the NCP can be constructed, the CI for RESI can be derived from the CI for NCP.
The confidence interval construction for NCP has been discussed for non-central Chi-squared and F distributions \citep{kent_confidence_1995, harlow_what_2013}. \\

Suppose the test statistics from the sample $T_{obs}^2$ is observed and its degrees of freedom is $m_1$. Let $F(T_{obs}^2; m_1, \lambda)$ denote the cumulative distribution function (CDF) of the non-central Chi-squared distribution the test statistic $T^2_{(o)}$ follows.
Then $F(T_{obs}^2; m_1, \lambda$) is a monotonic and strictly decreasing function of the NCP, $\lambda$. 

The lower bound ($\ell$) and upper bound ($u$) of central $(1 - \alpha)\times 100\%$ CI for $\lambda$ can be chosen as the values that satisfy the equalities \citep{kent_confidence_1995}
\begin{align*}
     F(T_{obs}^2; m_1, \lambda = \ell) = 1-\alpha/2\\
     F(T_{obs}^2; m_1, \lambda = u) = \alpha/2
\end{align*}

For the non-central F distribution, the procedure to construct a CI for its NCP is similar, except there are 2 degree of freedom parameters. The lower ($\ell$) and upper ($u$) bounds for the NCP of the F distribution can be determined from the observed statistic, $T_{obs}^2$, with degrees of freedom being $m_1$ and $(n-m)$ are chosen to satisfy
\begin{align*}
    F(T_{obs}^2; m_1, n-m, \lambda = \ell) = 1-\alpha/2\\
     F(T_{obs}^2; m_1, n-m, \lambda = u) = \alpha/2
\end{align*}

We provide code to compute these intervals in the {\tt RESI R} package (\url{https://github.com/statimagcoll/RESI}).
Our code is adapted from \citet{kent_confidence_1995} and \citet{harlow_what_2013}.

\subsubsection{Resampling methods: bootstrap CIs} 
As we discussed in the previous section, the distribution of the test statistic may deviate from the theoretical non-central Chi-squared or F distribution if the covariance must be estimated and/or the covariate(s) is random instead of fixed by designs. In this situation, the CIs built through non-central distributions may have bad coverage of the true effect size. Bootstrapping is a good alternative that can be used to approximate the actual distribution of the test statistic and can be used to construct the CIs for test statistic and the estimated effect size \citep{efron_bootstrap_1979,hall_bootstrap_1992}. 

In the body of this paper, we consider the standard nonparametric bootstrap by sampling the data with replacement $R$ times and estimating the corresponding RESI for each resampled data. The lower and upper $\alpha/2\times 100\%$ percentiles of the $R$ estimated RESIs are the bootstrapped lower and upper bounds of the $(1-\alpha)\times 100\%$ bootstrap CI for the estimated RESI. 

In addition to the non-parametric bootstrap, we also consider several variants of the wild bootstrap, which was originally proposed by \citet{wu_jackknife_1986}. The wild bootstrap is a kind of residual bootstrap which leaves the covariates at the sample value but resamples the outcome values based on the residual values. That is, in each replicate, the resampled outcome value of the $i$-th observation is
\begin{equation*}
    y_i^* = \hat{y}_i + \hat{\epsilon}_i \times w_i,
\end{equation*}
where $\hat{y}_i$ and $\hat{\epsilon}_i$ are the expected value and estimated residual of the $i$-th observation from the analysis model, respectively. $w_i$ is a random number, called multiplier, drawn from a distribution $W$ such that $\E(W) = 0$ and $Var(W) = 1$. 
There are several different distributions from which $w_i$ can be drawn, such as the Rademacher and standard Normal distributions. 

The original wild bootstrap doesn't resample the data. We propose several variants of the original wild bootstrap: (1) resampling covariates along with residuals with replacements; (2) fixing covariates and only resampling residuals with replacements; (3) independently resampling covariates and residuals with replacements. We also considered 3 types of multipliers: (1) no multipliers, i.e., $w_i = 1$, $\forall i$; (2) multipliers drawn from the Rademacher distribution; (3) multipliers drawn from standard Normal distribution. Table \ref{tab: boot_summary} summarizes the original and variant wild bootstraps considered in this paper. The original wild bootstrap without using multipliers is excluded because no resampling happens within this combination. Therefore, there are 11 combinations of the wild bootstrapping evaluated in the simulations. Detailed results for all the bootstrap methods are provided in the Supplementary Material.

\begin{table}
\tiny
    \centering
    \begin{tabular}[t]{p{3 cm} l l p{4.5cm}}
    \toprule
        Wild bootstrap type & Multiplier & Bootstrap model & Assumption(s)  \\
    \midrule
        Original & None & $Y^{(b)} = X\hat{\beta} + \hat{\epsilon} \times 1  = Y$ & Excluded b/c no sampling happens here\\
                                & Rademacher & $Y^{(b)} = X\hat{\beta} + \hat{\epsilon} \times W$,  $w_i \sim Rad$  & Symmetric errors\\
                                & $N(0, 1)$ & $Y^{(b)} = X\hat{\beta} + \hat{\epsilon} \times W$, $w_i \sim N(0, 1)$  & Symmetric errors\\
                                 \cline{2-4}
                                
         (1) Resampling covariates along with residuals 
         & None & $Y^{(b)} = R^{(b)}X\hat{\beta} + R^{(b)}\hat{\epsilon} \times 1$  & Non-parametric bootstrap. not assuming homo-/hetero-skedasticity or symmetric errors\\
         & Rademacher & $Y^{(b)} = R^{(b)}X\hat{\beta} + R^{(b)}\hat{\epsilon} \times W$, $w_i \sim Rad$ & Symmetric errors\\
         & $N(0, 1)$ & $Y^{(b)} = R^{(b)}X\hat{\beta} + R^{(b)}\hat{\epsilon} \times W$, $w_i \sim N(0, 1)$ & Symmetric errors\\
         \cline{2-4}
         (2) Fixing covariates and only resampling residuals 
         & None & $Y^{(b)} = X\hat{\beta} + R^{(b)}\hat{\epsilon} \times 1$  & Homoskedasticity \\
         & Rademacher & $Y^{(b)} = X\hat{\beta} + R^{(b)}\hat{\epsilon} \times W$, $w_i \sim Rad$ & Homoskedasticity and symmetric errors\\
         & $N(0, 1)$ & $Y^{(b)} = X\hat{\beta} + R^{(b)}\hat{\epsilon} \times W$, $w_i \sim N(0, 1)$& Homoskedasticity and symmetric errors\\
         \cline{2-4}
         (3) independently resampling covariates and residuals 
         & None & $Y^{(b)} = R^{(b)}_1 X\hat{\beta} + R^{(b)}_2 \hat{\epsilon} \times 1$  & Homoskedasticity \\
         & Rademacher & $Y^{(b)} = R^{(b)}_1 X\hat{\beta} + R^{(b)}_2 \hat{\epsilon} \times W$, $w_i \sim Rad$ & Homoskedasticity and symmetric errors\\
         & $N(0, 1)$ & $Y^{(b)} = R^{(b)}_1 X\hat{\beta} + R^{(b)}_2 \hat{\epsilon} \times W$, $w_i \sim N(0, 1)$ & Homoskedasticity and symmetric errors\\
         \cline{2-4}
    \toprule                                                   
    \end{tabular}
    \caption{Eleven bootstrap procedures considered in this paper. The original wild bootstrap with constant multiplier is excluded since it performs no resampling.
    $R^{(b)}$ denotes the bootstrap matrix for the $b$-th replicate; $Y^{(b)}$ is the bootstrapped outcome values; $X$ is the covariate(s); $\hat{\beta}$ is the estimated parameters; $\hat{\epsilon}$ is the estimated residuals; $W$ is the randomly drawn values for the multiplier.}
    \label{tab: boot_summary}
\end{table}

\subsection{Simulation setup} \label{sec: sim_1}
In the previous sections, we proposed three estimators and several ways of constructing CIs for RESI. In this section, we use 1,000 simulations to evaluate the performance of the proposed CI construction procedures with respect to each estimator under 128 different scenarios.
We evaluate the influence of heteroskedasticity, data skewness and fixed/random covariate(s) on the performance of estimators and CIs.
All CIs are constructed at the significance level of 0.05.\\

% For simplicity, we begin our simulation studies by considering a situation where there is only one target covariate and no nuisance covariate.

We simulate a simple linear regression model $Y = \beta_0 + \beta_1 x + \epsilon$, where $x \in \{0, 1\}$.
We vary the sample size $n \in \{ 50, 100, 250, 500\}$  across 4 different true effect sizes, $S \in \{0, 0.33, 0.66, 1\}$.
In the scenario of symmetric errors, the errors are independently sampled from $N(0, \sigma_0^2 = 1^2)$ under homoskedasticity;
Under heteroskedasticity, the errors are independently sample from $N(0, \sigma_0^2 = 0.5^2)$ if $x_i = 0$ and from $N(0, \sigma_1^2 = 1.5^2)$ if $x_i = 1$.
In the scenario of skewed errors, $\epsilon + \sqrt{0.1}$ $\sim$ Gamma($\alpha = 0.1$, $\beta = \sqrt{0.1}$) under homoskedasticity and $\epsilon + \sqrt{0.1}\cdot(x+0.5)$ $\sim$ Gamma($\alpha = 0.1$, $\beta = \sqrt{0.1}/(x+0.5)$) under heteroskedasticity. These errors are very heavily right-skewed, with Pearson's moment coefficient of skewness equal to $2/\sqrt{0.1} \approx 6.32$.
To illustrate the difference in the performance of CI between a randomized controlled and observational design where the covariate(s) $X$ is treated as fixed or random, the values of the covariate were generated in two ways: (1) when the covariate is fixed, $\pi = 0.3$ and ceiling($n\pi$) of the $n$ individuals have their covariate with value 1 and the remaining have value 0; (2) when the covariate is random, $X \sim \mathrm{Bern}(0.3)$ is sampled from a Bernoulli distribution with parameter $\pi = 0.3$. 
For each bootstrap CI, 1,000 bootstraps are used.

% \subsubsection{Simulation 2: with nuisance covariates} \label{sec: sim_2}
% blah blah blah

\section{Simulation Results} \label{sec: sim_results}

We ran simulations to assess the bias of the estimators and the coverage of the confidence intervals.
We considered eight possible cases where there are homo- or hetero-skedaticity errors, symmetric or skewed errors, and fixed or random covariate in each simulation.

\begin{figure}[h]
    \centering
    \includegraphics[width = 16cm]{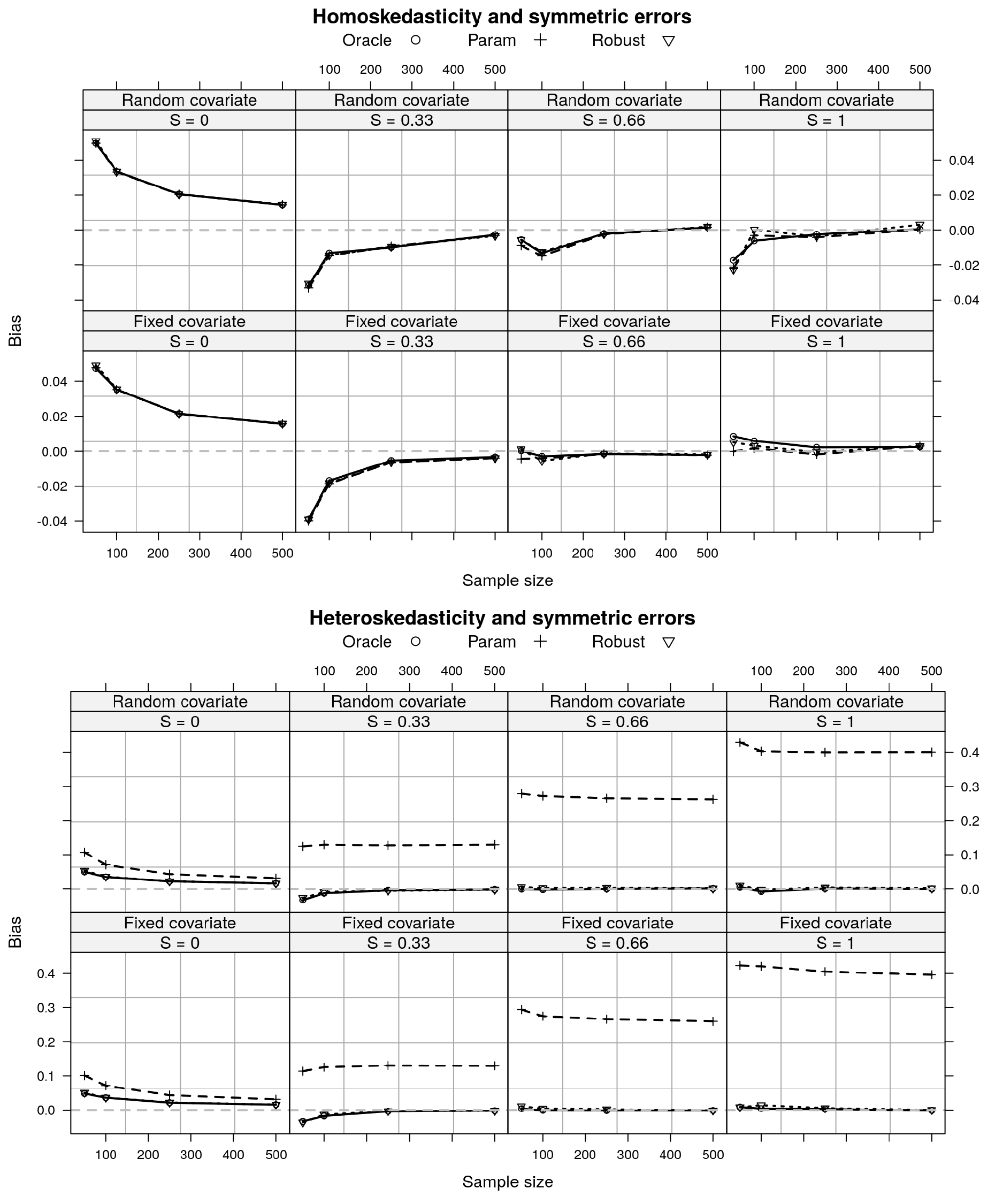}
    \caption{A comparison of bias for three estimators of the RESI under the simulation settings described in Section \ref{sec: sim_1}. Under homoskedasticity the estimators have small bias (top). Under heteroskedasticity, the parametric estimator for effect size is heavily biased. $S$ denotes the RESI, Oracle estimator assumes known variance, parametric assumes homoskedasticity, Robust is the sandwich covariance estimator (Section \ref{sec: linear_case}).}
    \label{fig:bias1}
\end{figure}

\begin{figure}[h]
    \centering
    \includegraphics[width = 16cm]{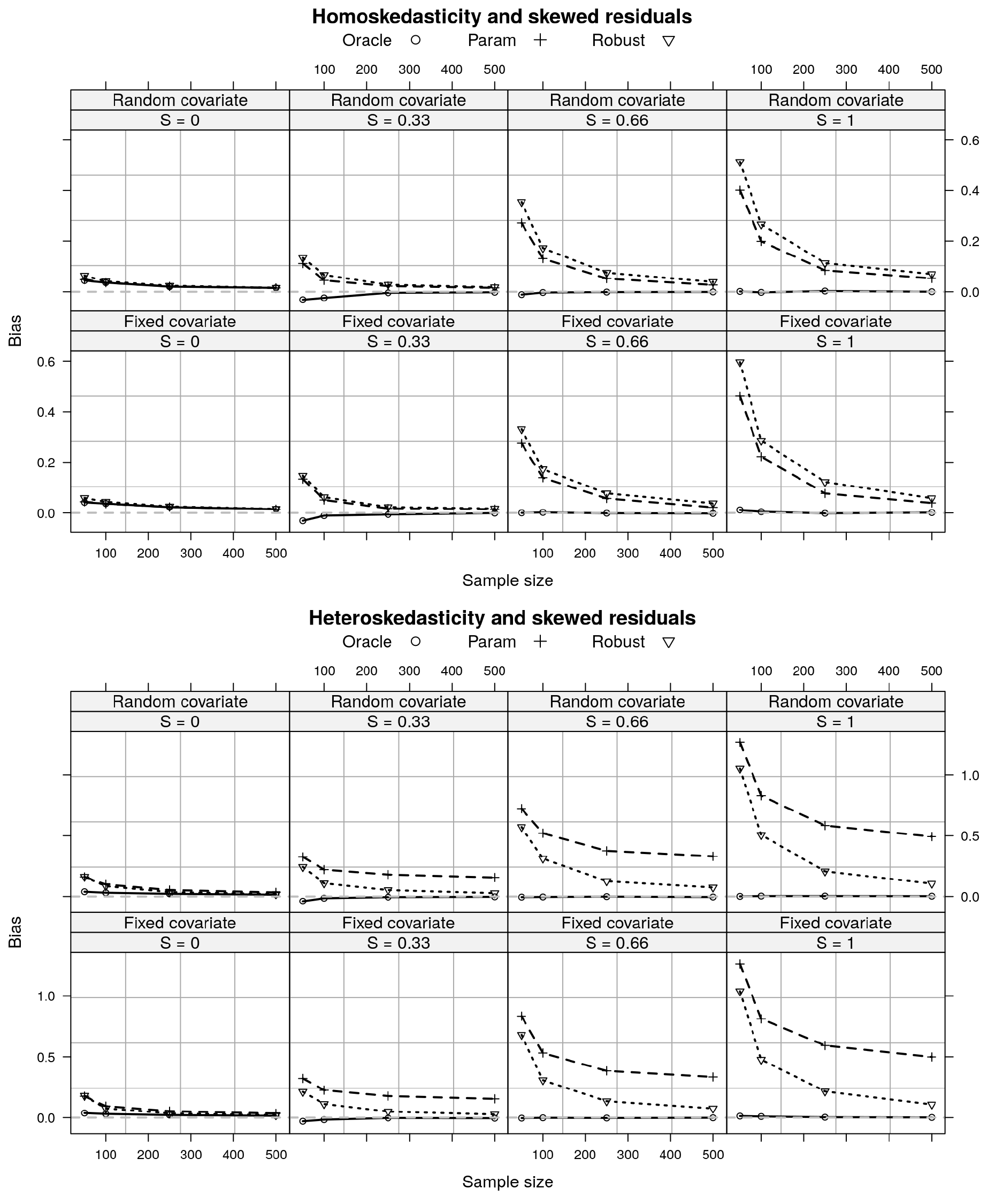}
    \caption{A comparison of bias for three estimators of the RESI under the simulation settings described in Section \ref{sec: sim_1}. Under homoskedasticity and heavily skewed residuals, the estimators have bias (top), but the bias goes to 0 in large samples. Under heteroskedasticity and heavily skewed residuals, the parametric estimator for effect size is heavily biased, the bias of the robust estimator approaches 0 in large samples. $S$ denotes the RESI, Oracle estimator assumes known variance, parametric assumes homoskedasticity, Robust is the sandwich covariance estimator (Section \ref{sec: linear_case}).}
    \label{fig:bias2}
\end{figure}

% Bias
In small samples ($n=50$) the estimators are positively biased for $S=0$, but negatively biased for all other values of $S$ (Figure \ref{fig:bias1}).
For all other sample sizes $\hat{S}$ has small bias.
As expected, under heteroskedasticity, the parametric estimator is heavily biased \citep{vandekar_robust_2020}.
When the error distribution is heavily skewed and  $S>0$, the robust estimator is biased, but the bias goes to zero in large samples (Figure \ref{fig:bias2}).
The randomness of covariate doesn't have an effect on the consistency of all these 3 estimators.

\begin{figure}[h]
    \centering
    \includegraphics[width = 14cm]{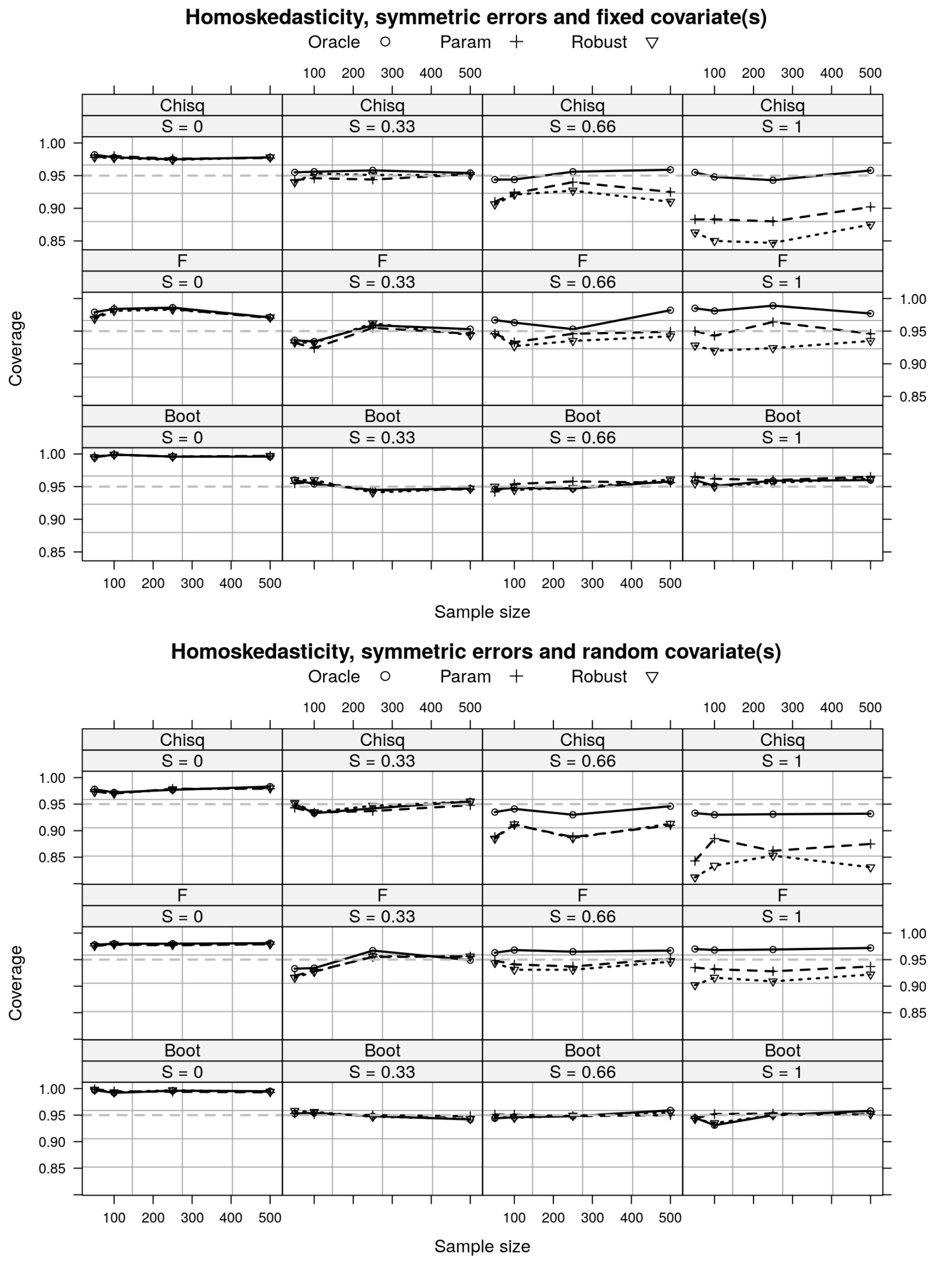}
    \caption{A comparison of the coverage of different CIs for the three estimators of RESI under homoskedasticity, with symmetric residuals and random/fixed covariates. With fixed covariates, the Chi-squared CI has nominal coverage for the oracle estimator and F CI has nominal coverage for the parametric estimator, whereas they both fail when the covariates are random. The bootstrapped CI has nominal coverage for all estimators, with random or fixed covariates. 
    $S$ denotes the RESI, Oracle estimator assumes known variance, parametric assumes homoskedasticity, Robust is the sandwich covariance estimator (Section \ref{sec: linear_case})
    }
    \label{fig:converage_1}
\end{figure}

% Coverage: the influence of fixed/random covariate.
The pattern of differences between the oracle, parametric, and robust estimators is most obvious when the effect size is larger (e.g. $S=1$) with fixed covariate (Figure \ref{fig:converage_1}).
In this case, the Chi-squared CI should have nominal coverage for the oracle estimator, the F CI should have nominal coverage for the parametric estimator and the bootstrap CI should should have nominal coverage for all (asymptotically; Figure \ref{fig:converage_1}).
As expected, larger effect sizes have worse coverage for the Chi-squared and F CIs when the variance is estimated (parametric and robust statistics; Figure \ref{fig:converage_1}) because the variance depends on the true effect size (see Section \ref{sec: linear_case}).

% The impact of fixed/random covariate
When there is random covariate, the Chi-squared and F CIs both fail to provide nominal coverage for the oracle and parametric estimators, respectively. This is because the extra variance introduced by random covariate(s) into the test statistics (also shown in Figure \ref{fig: var_of_stat}).  

\begin{figure}[h]
    \centering
    \includegraphics[width = 14cm]{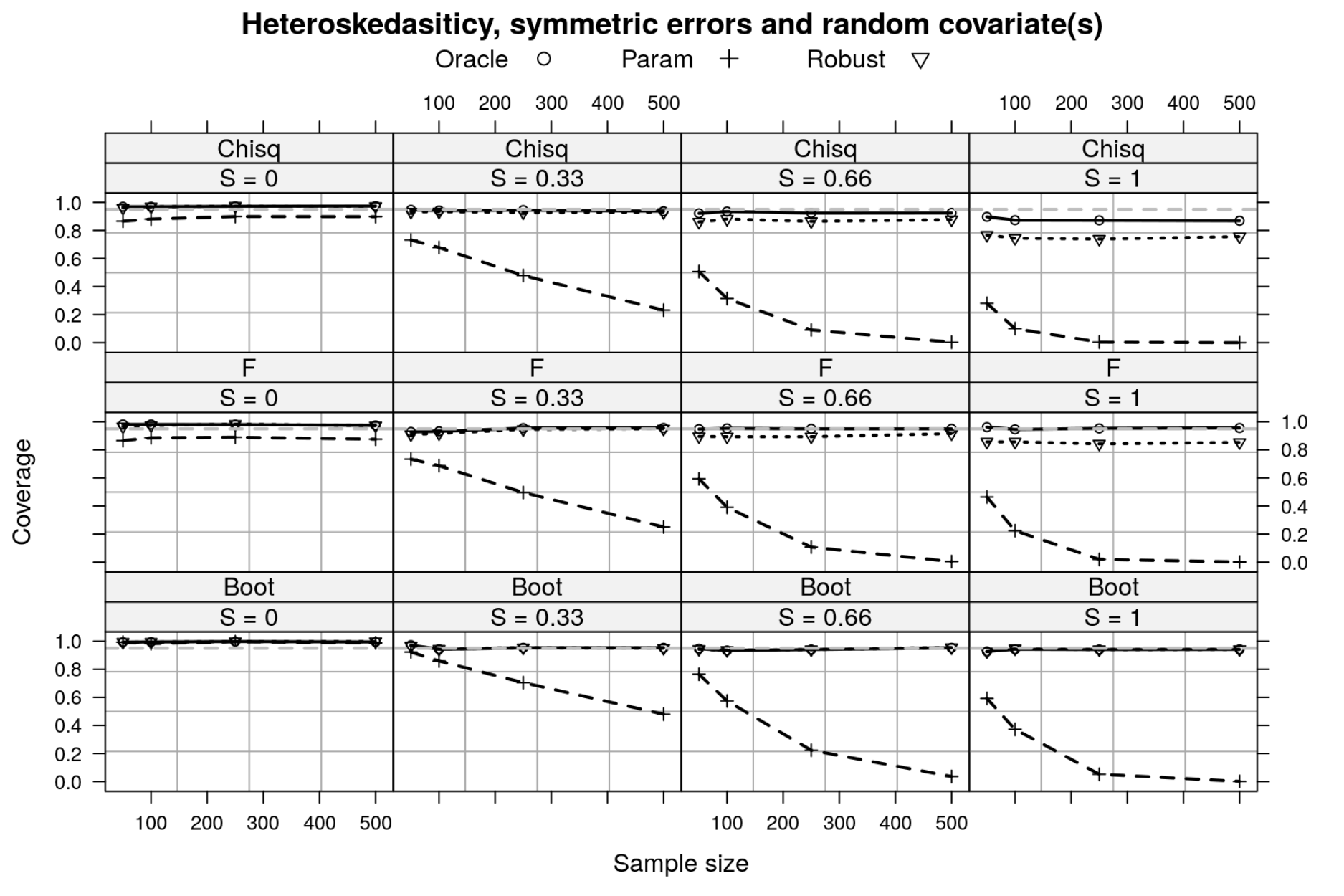}
    \caption{A comparison of the coverage of different CIs for the three estimators of RESI under heteroskedasticity, with symmetric residuals and random covariates. All three CIs fail to provide nomial coverage for the parametric estimator. The Chi-squred and F CIs fail to provide nominal coverage for the robust estimator. The bootstrapped CI has nominal coverage for both the oracle and robust estimators.
    $S$ denotes the RESI, Oracle estimator assumes known variance, parametric assumes homoskedasticity, Robust is the sandwich covariance estimator (Section \ref{sec: linear_case})
    }
    \label{fig:converage_2}
\end{figure}

% Coverage: the influence of heteroskedasticity.
Under heteroskedasticity, all CIs fail to provide nominal coverage for the parametric estimator as expected because this estimator is biased as shown in Figure \ref{fig:bias1}. Both Chi-squared and F CIs fail to provide nominal coverage for the robust estimator. The bootstrap CI has nominal coverage for both of the oracle and robust estimators (Figure \ref{fig:converage_2}). 

\begin{figure}[h]
    \centering
    \includegraphics[width = 14cm]{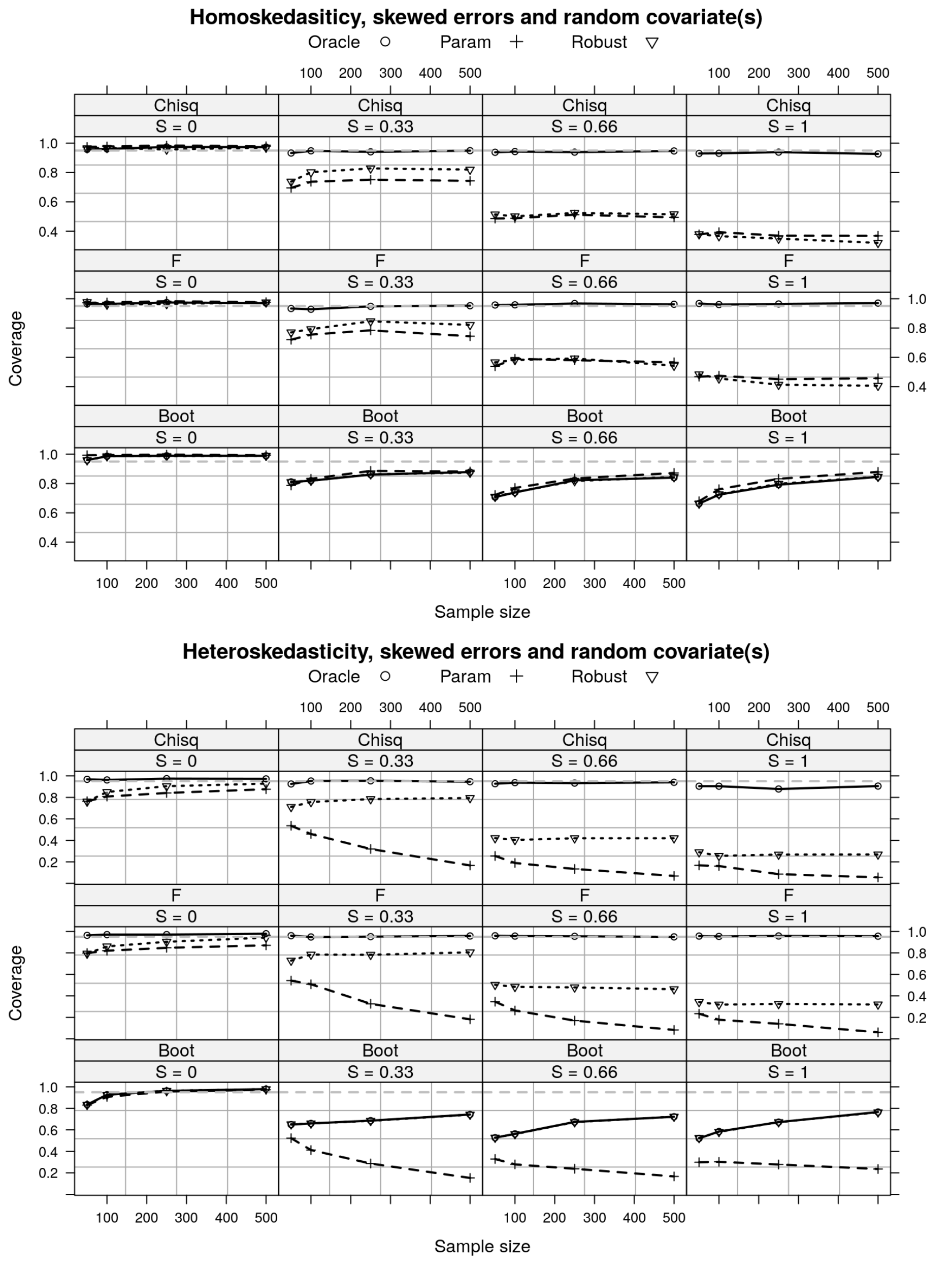}
    \caption{A comparison of the coverage of different CIs for the three estimators of RESI with skewed residuals and random covariates. 
    In homoskedasticity, the coverage of the bootstrapped CI approaches to the nominal level in large sample for all three estimators. In heteroskedasticity, the boostrapped CI approaches to the nominal level in large samples for the oracle and robust estimators.
    $S$ denotes the RESI, Oracle estimator assumes known variance, parametric assumes homoskedasticity, Robust is the sandwich covariance estimator (Section \ref{sec: linear_case})}
    \label{fig:converage_3}
\end{figure}

% Coverage: the influence of heteroskedasticity in the existence of skewed residuals.
The skewness of the residuals has a big impact on the CIs' performance. 
Under homoskedasticity, the Chi-squared and F CIs provide nominal coverage for the oracle estimator, but they both fail to provide nominal coverage for the parametric and robust estimators (Figure \ref{fig:converage_3}). In large samples, the coverage of the bootstrap CI for all three estimators approaches to the nominal level. 
Under heteroskedasticity, all CIs fail to provide nominal coverage for all estimators except F CI for the oracle estimator. The coverage of bootstrap CI for the oracle and robust estimators approaches to the nominal level in large samples. 

\section{Application of analysis of effect sizes} \label{sec: data_analysis}

In this section, we use two datasets from studies of relational memory among schizophrenia and early psychosis patients \citep{armstrong_impaired_2012, avery_relational_2021} to illustrate how to conduct the analysis of effect sizes (ANOES) with the function {\tt anoes} in the {\tt RESI R} package. 
Relational memory is the ability to bind information into complex memories and is impaired in chronic schizophrenia and in the early stages of psychosis \citep{armstrong_impaired_2012, armstrong_impaired_2018, avery_relational_2021}. In both studies a relational memory paradigm was used to compare the ability of psychosis patients and healthy individuals to identify novel stimulus pairings \citep{armstrong_impaired_2012, armstrong_impaired_2018, avery_relational_2021}. 
The first study compared relational memory accuracy in 60 patients with schizophrenia or schizoaffective disorder to 38 healthy control subjects \citet{armstrong_impaired_2012}. 
The second study assessed participant's relational memory accuracy in 66 early psychosis patients and 64 healthy control subjects \citep{avery_relational_2021}.

In our analyses, in order to demonstrate the communicability of the RESI across different models, we use both logistic regression models and multiple linear models to quantify the differences in relational memory performance in schizophrenia and early psychosis after controlling for age and gender, respectively (See Appendix for {\tt R} code). Then, ANOESs are conducted based on the model results as the RESI and corresponding CIs are estimated to indicate the effect size of each factor after controlling others. The results are summarized in an ANOVA table format (see Table \ref{tab: data_analysis_SZ} and \ref{tab: data_analysis_EP}).
We perform the ANOES on the model fit object using the {\tt anoes} function in {\tt R} with the robust covariance estimator, and the nonparametric bootstrap with 1,000 bootstrap samples. We include estimates of effect size and CIs for each parameter estimate as well as for the overall model fit.
We also include the results using the hypothesis testing framework output by the analysis of variance.

% By default, it calculates the RESI for each individual factor (after controlling for others) as well as for all factors together (by comparing with the reduced model which only contains the intercept, i.e., {\tt glm(accuracy $\sim$ 1, data = data, family = "binomial")}), and it also estimates the CIs via non-parametric bootstraps for each RESI estimates at the pre-specified significance level of {\tt alpha = 0.05}. 
% It uses the robust (sandwich) estimator ({\tt robust.var = TRUE}, default) for the RESI and 1,000 bootstraps ({\tt nboot = 1000}, default) to construct the CIs. 

Table \ref{tab: data_analysis_SZ} shows that the estimated RESI of schizophrenia (after controlling for age and gender) is 0.44 with 95\% CI (0.24, 0.63) from the logistic model and 0.46 with 95\% CI (0.24, 0.69) from the linear model. 
Table \ref{tab: data_analysis_EP} shows that the estimated RESI of early psychosis (after controlling for age and gender) is 0.49 with 95\% CI (0.31, 0.72) from the logistic model and 0.50 with 95\% CI (0.32, 0.71) from linear model. 

The estimated RESIs and their CIs from logistic and linear models are very close to each other.
If one model had much greater sensitivity to a given effect, it would be represented by a larger effect size comparing across the models.
Using the RESI makes it easier to compare across logistic and linear models.

The RESI also makes it easy to compare findings across these two studies that have different sample sizes.
While the test statistics are sample size dependent, the RESI can be compared across the four models that come from two different samples and two model types.

%to study the same topics like ours (the effect of schizophrenia/early psychosis on the accuracy of recognizing premise pairs), as they would use different effect size indices (such as standardized log odds ratio for the logistic models and $R^2$ for the linear model). Whereas now, with RESI they can conveniently communicate their observed effect sizes through different types of models.

\begin{table}[ht]
  \centering
  \begin{tabular}{l c c c c c c c }
    \multicolumn{8}{l}{\textbf{Logistic model: schizophrenia (SZ) vs. healthy controls}} \\
    \toprule
    factor & estimate & robust s.e. & Chi-squared & d.f. & $p$-value & RESI & 95\% CI\\
    \midrule
    group (SZ)      & -0.77 & 0.18 & 18.92 & 1 & $<$0.001 & 0.44 & (0.25, 0.63)\\ 
    age             & -0.03 & 0.01 & 14.18 & 1 & $<$0.001 & 0.37 & (0.17, 0.59)\\
    gender (female) & -0.15 & 0.14 & 1.19  & 1 & 0.27     & 0.05 & (0, 0.30)\\
    \midrule
    \textbf{Overall} & & & 50.55 & 3 & $<$0.001 & 0.71 & (0.53, 1.02)\\
    Residual         & & & & 94 & & & \\
    \bottomrule
    \\
    \multicolumn{8}{l}{\textbf{Linear model: schizophrenia (SZ) vs. healthy controls}} \\
    \toprule
    factor & estimate & robust s.e. & Chi-squared & d.f. & $p$-value & RESI & 95\% CI\\
    \midrule
    group (SZ)      & -0.17 & 0.04 & 20.99 & 1 & $<$0.001 & 0.46 & (0.24, 0.68)\\ 
    age             & -0.01 & 0.00 & 15.35 & 1 & $<$0.001 & 0.39 & (0.18, 0.65)\\
    gender (female) & -0.03 & 0.03 & 1.00  & 1 & 0.32    & 0    & (0, 0.29) \\
    \midrule
    \textbf{Overall} & & & 72.76 & 3  & $<$0.001 & 0.86 & (0.60, 1.38)\\
    Residual         & & & & 94 & &  & \\
    \bottomrule

  \end{tabular}
  \caption{A comparison of ANOESs of the schizophrenia (SZ) from logistic and linear models. The observed RESIs and their CIs for SZ after controlling for age and gender are very close from the two different models, which demonstrates the robustness of RESI across different models. The $p$-values are calculated from the Chi-squared distribution.}
  \label{tab: data_analysis_SZ}
\end{table}

\begin{table}[ht]
  \centering
  \begin{tabular}{l c c c c c c c }
    \multicolumn{8}{l}{\textbf{Logistic model: early psychosis (EP) vs. healthy controls}} \\
    \toprule
    factor & estimate & robust s.e. & Chi-squared & d.f. & $p$-value & RESI &  95\% CI\\
    \midrule
    group (EP)      & -1.26 & 0.22 & 31.43 & 1 & $<$0.001 & 0.49  & (0.31, 0.72)\\
    age             & 0.04  & 0.04 & 0.79  & 1 & 0.37    & 0     & (0, 0.28)\\
    gender (female) & 0.33  & 0.27 & 1.50  & 1 & 0.22    & 0.06  & (0, 0.26)\\
    \midrule
    \textbf{Overall} & &   & 45.17 & 3   & $<$0.001 & 0.58 & (0.43, 0.82)\\
    Residual         & &   &       & 126 &  &  & \\
    \bottomrule
    \\
    \multicolumn{8}{l}{\textbf{Linear model: early psychosis (EP) vs. healthy controls}} \\
    \toprule
    factor & estimate & robust s.e. & Chi-squared & d.f. & $p$-value & RESI &  95\% CI\\
    \midrule
    group (EP)      & -0.18 & 0.03 & 32.54 & 1 & $<$0.001 & 0.50  & (0.32, 0.71)\\
    age             & 0.01  & 0.01 & 0.93  & 1 & 0.33    & 0      & (0, 0.29)\\
    gender (female) & 0.04  & 0.03 & 1.73  & 1 & 0.19    & 0.08   & (0, 0.29)\\
    \midrule
    \textbf{Overall} & & & 45.59 & 3 & $<$0.001  & 0.58 & (0.44, 0.82)\\
    Residual         & & & & 126 &  &  & \\
    \bottomrule
  \end{tabular}
  \caption{A comparison of ANOESs of the early psychosis (EP) from logistic and linear models. The observed RESIs and their CIs for EP after controlling for age and gender are very close from the two different models, which demonstrates the robustness of RESI across different models. The $p$-values are calculated from the Chi-squared distribution.}  
  \label{tab: data_analysis_EP}
\end{table}

\section{Discussion}
In this paper, we derived confidence intervals (CIs) for the robust effect size index (RESI) and used them to describe an analysis of effect sizes (ANOES) approach.
% overview of important stuff in this paper.
We proposed 3 different estimators for the RESI and a variety of ways to construct CIs. The oracle estimator assumes the covariance matrix is known, so is not possible to compute in applications.
% more details.
Through simulations, we showed that all 3 estimators are consistent under homoskedasticity and the robust estimator is consistent under heteroskedasticity. 
We used statistical theory and simulations to demonstrate that the non-central Chi-squared CI has low coverage when the covariance in the Wald test statistic needs to be estimated.
In addition, the randomness of covariate(s) also reduces the coverage of Chi-squared and F CIs, which is an important implication for the observational studies where the covariates are random instead of fixed/controlled by the experimenters.
According to the simulation results, using the robust estimator along with the (non-parametric) bootstrap CI is generally most accurate and applicable to conduct consistent estimation and valid inference using the RESI. 
% The result of real data analysis.

% why is this work important
%In the past, many researchers might be too cautious to engage too deeply with effect sizes because of their lack of familiarity with them, let alone they had to properly select the effect sizes suitable for their models, correctly estimate them and calculate accurate CIs for them. The lack of guidance for CI construction for effect sizes even exacerbates this difficulty.
The RESI estimator and CI reduce the barriers to effect size analysis by introducing a single framework that is widely applicable across different models.
They provide a basis for a framework of the ANOES, where effect sizes with confidence intervals can easily be reported in summary tables alongside $p$-values. 
This approach may help to address the limitations of null hypothesis significance testing \citep{wasserstein_asas_2016,wasserstein_moving_2019} and may provide guidance on whether a study finding is under-powered. 
We hope the ease of using RESI will broaden the use of effect sizes and their CIs in study reporting.

We also provided functions to perform ANOES based on the RESI in our {\tt RESI R} package (\url{https://github.com/statimagcoll/RESI}). 
It outputs the estimated RESI along with CIs in a summary table, with which researchers can conveniently report their observed effect sizes along with trustworthy CIs.
Coupled with the generality of the RESI, researchers studying the same scientific questions but using different types of data can easily communicate their observed effect sizes and CIs instead of having to translate between different effect size indices.
 
% LIMITATIONS
Our research focused on the performance of the CIs in a linear regression model setting because these are widely used models. Although it adequately illustrated the problems we wanted to discuss, theory and evaluation for other models (e.g., mixed effects models and survival analysis) requires further research.
% Future direction
We formed the basis of applying the RESI in the ANOES on cross-sectional data. It can be expected that there would be more technical questions to be solved before making solid ANOES in longitudinal studies. In the future, it would be interesting to work on the estimation and inference of the RESI with longitudinal data.

\newpage

\bibliographystyle{apalike}
\bibliography{./MyLibrary}

\section*{Appendix}

\subsection*{ANOES R code}

In this section, we illustrate how to perform an ANOES in the relational memory datasets using the functions from the {\tt RESI R} package, and the complete code for our data analyses are also included. 
The two datasets are described in section \ref{sec: data_analysis} and have already been loaded into {\tt R}. Each of these two datasets contains 4 variables ``accuracy" (the outcome), ``group", ``age" and ``gender". 
We first fit a {\tt glm} object by specifying the variable {\tt accuracy} as the outcome, it can be a logistic regression model if we specify {\tt family = "binomial"} or a linear model if {\tt family = "gaussian"} (default). 
Note, for both models, the assumptions are violated because, for the logistic regression the data are not Bernoulli distributed (they are proportions) and for the Gaussian model, the errors are not normal.
Then the ANOES based on the RESI can be implemented with {\tt anoes(model.full = glm.fit)}, where {\tt glm.fit} is the glm object we just created. 

The complete codes for our analyses are:

\begin{lstlisting}[language=R]
# ------------------------------------------- #
# ANOES in the relational memory data #
# ------------------------------------------- #
# Install the package RESI from Github
devtools::install_github("https://github.com/statimagcoll/RESI")
# Load in package(s)
library(RESI)
## The two datasets have already been loaded, called "SZ" and "EP"
## Each of them contains 3 variables called "accuracy", "age" and "gender"

## Set seed number (the `anoes()` function uses bootstraps to build CIs by default)
set.seed(1213)

# Schizophrenia (SZ) patients vs Healthy control (HC) subjects
## fit a logistic model using `glm()`
SZ_log_mod <- glm(accuracy ~ group + age + gender, data = SZ, family = "binomial")
## conduct ANOES based on the logistic model
anoes(model.full = SZ_log_mod)
## fit a mutiple linear model
SZ_lin_mod <- glm(accuracy ~ group + age + gender, data = SZ, family = "gaussian")
## conduct ANOES based on the linear model
anoes(model.full = SZ_lin_mod)

# Early psychosis (EP) patients vs healthy control (HC) subjects
## fit a logistic model using `glm()`
EP_log_mod <- glm(accuracy ~ group + age + gender, data = EP, family = "binomial")
## conduct ANOES based on the logistic model
anoes(model.full = EP_log_mod)
## fit a mutiple linear model
EP_lin_mod <- glm(accuracy ~ group + age + gender, data = EP, family = "gaussian")
## conduct ANOES based on the linear model
anoes(model.full = EP_lin_mod)
\end{lstlisting}

By default, {\tt anoes} uses the robust (sandwich) estimator ({\tt robust.var = TRUE}, default) to estimate the RESI, if {\tt robust.var = FALSE}, it uses \eqref{eq:Jestimate} to calculate the parametric test statistics \eqref{eq: para_test_stat} and corresponding parametric version of RESI estimates. 
{\tt anoes} uses 1,000 bootstraps ({\tt nboot = 1000}) to construct the corresponding CIs. 
{\tt alpha} is the significance level at which the CIs are estimated, by default, {\tt alpha = 0.05}; 

If we are interested in estimating the overall effect size of a subset of factors (e.g., age and gender in our example), we can specify a reduced model to compare with (e.g., {\tt model.reduced = glm(accuracy $\sim$ group, data = data, family = "binomial")}) instead. By default, {\tt model.reduced = NULL}. The function will then output a table containing the overall effect size of age and gender after controlling for the group. Here is an example:
\begin{lstlisting}
> mod1 = glm(accuracy ~ group + age + gender, data = SZ, family = "gaussian")
> mod0 = glm(accuracy ~ group, data = SZ, family = "gaussian")
> anoes(model.full = mod1, model.reduced = mod0)
         Chi-squared df p-val  RESI    LL   UL
Tested         16.99  2     0 0.399 0.207 0.72
Residual          NA 94    NA    NA    NA   NA
\end{lstlisting}
where the RESI on the ``tested" row is the RESI estimate for age and gender together after controlling for group.

\subsection*{Mathematical details}

To show \eqref{eq:F_dist}: the distribution is a non-central F distribution.\\
    Under the homoskedasticity and Normality assumption, we have $Y \sim N(X\beta, \sigma^2 \mathbf{I})$.\\
    \begin{align*}
        T_{(p)}^2 / m_1 & = n \hat{\sigma}^{-2} (\hat{\beta}_{OLS} - \beta_0)^T X^T X (\hat{\beta}_{OLS} - \beta_0) /  m_1 \\
        & = \frac{ \sigma^{-2} n(\hat{\beta}_{OLS} - \beta_0)^T X^T  X (\hat{\beta}_{OLS} - \beta_0) / m_1 }{\sigma^{-2} \hat{\sigma^2}} \\
        & = \frac{\sigma^{-2} n(\hat{\beta}_{OLS} - \beta_0)^T X^T  X (\hat{\beta}_{OLS} - \beta_0)}{\sigma^{-2} Y^T(I - H)Y /  (n - m) }
    \end{align*}
    \begin{enumerate}
        \item In the numerator, under mild conditions, the quadratic form $\sigma^{-2} n(\hat{\beta}_{OLS} - \beta_0)^T X^T  X (\hat{\beta}_{OLS} - \beta_0) \sim \chi^2(m_1; \lambda_1)$, where $\lambda_1$ is the non-centrality parameter:
        \begin{align*}
            \lambda_1 & = \sigma^{-2} n(\E\hat{\beta}_{OLS} - \beta_0)^T X^T  X (\E\hat{\beta}_{OLS} - \beta_0) \\
            & = \sigma^{-2} n(\beta - \beta_0)^T X^T  X (\beta- \beta_0) \\
            & = nS_{\beta}^2
        \end{align*}
        This quadratic form in the numerator is essentially the regression sum of squares (SSR) in linear regression models.
        \item In the denominator, the quadratic from $\sigma^{-2} Y^T(I - H)Y \sim \chi^2(n-m; \lambda_2)$, where $\lambda_2$ is the non-centrality parameter. It can be shown that $\lambda_2$ is 0:
    \begin{align*}
        \lambda_2 & = \sigma^{-2} \left[ (X\beta)^T (I - H) X\beta \right] \\
        & = \sigma^{-2} \left[ \beta^T X^T X\beta - \beta^T X^T H X\beta \right] \\
        & = \sigma^{-2} \left[ \beta^T X^T X\beta - \beta^T X^T X (X^T X)^{-1} X^T X\beta \right] \\
        & = \sigma^{-2} \left[ \beta^T X^T X\beta - \beta^T X^T X \beta \right] \\
        & = 0
    \end{align*}
    Therefore, in the denominator, $\sigma^{-2}Y^T(I - H)Y$ follows a central Chi-squared distribution $\chi^2_{n-m}$. This quadratic form is essentially the error sum of squares (SSE) in linear regression models.\\
    \item Under homoskedasticity and Normality, the SSR and SSE are independent. Therefore, we can prove that the distribution of $T_{(p)}^2 / m_1$ is the non-central F distribution $F(m_1, n-m; nS^2_{\beta})$
    \end{enumerate}

\end{document}